\documentclass[12pt]{article}
\usepackage{amsmath,amsthm,amsfonts}
\usepackage{cite}
\theoremstyle{definition}

\theoremstyle{definition}

\theoremstyle{definition}
 
\begin{document}
\pdfoutput=1 
{\LARGE \bf 
Symmetry  of Lie algebras associated with
$(\varepsilon,\delta )$ Freudenthal-Kantor triple systems}
\begin{center}
{\Large Noriaki Kamiya$^{1}$and Susumu Okubo$^{2}$}
 \end{center}
 \par
 \noindent
 $^{1}$Department of Mathematics, University of Aizu\\
 Aizuwakamatsu, JAPAN, kamiya@u-aizu.ac.jp \\
$^{2}$Department of Physics and Astronomy, University of Rochester \\
Rochester, NY 14627 U.S.A, okubo@pas.rochester.edu \\
\date{}
%
\vskip 3mm
\begin{abstract}
Symmetry group of Lie algebras and superalgebras constructed from
$(\epsilon,\delta)$ Freudenthal-Kantor 
triple systems has been studied.
Especially, for a special 
$(\varepsilon,\varepsilon)$ Freudenthal-Kantor triple,
it is $SL(2)$ group.
Also,
relationship between two 
constructions of Lie algebras from 
structurable algebra has been investigated.
\vskip 3mm
\noindent
AMS classification: 17C50; 17A40; 17B60 \\
Keywords:triple systems, 
Lie algebras, symmety group.\\
\end{abstract}
\par
\vskip 3mm
{\bf 1.\ Introduction and Summary of Main Results}
\vskip 3mm
Let $L$ be a Lie algebra over a field $F$.
Suppose that it is endowed with a group 
homomorphism
$$
G \rightarrow  Aut_{G}(L)
$$ 
for a group $G$.
If $G$ is a finitely generated Abelian group, 
a grading of $L$ is given (see [Ko]) 
by the action of the automorphism
(the group of characters) 
on the Lie algebras.
Recently,
we have studied actions of 
some small finite-dimensional 
non-Abelian groups by
automorphism on a 
Lie algebra $L$.
For instance, 
Lie algebras with symmetric group $S_{4}$ as its 
automorphism are
coordinated
([E-O.1]) 
by some non-associative algebras.
The unital algebras in this class
turn out to be the 
structurable algebras of 
Allison ([A]).
Moreover,
Lie algebras graded by 
the non-reduced root system $BC_{1}$
of type $B_{1}$ 
([B-S.])
are naturally among 
the Lie algebras with the
$S_{4}$-symmetry.
We have also studied 
the case of the invariant group
$G$ being 
$S_{3}$
and more generally dicyclic 
group $Dic_{3}$
([E-O.2,3])
for the characterization 
of $L$. 
\par
Since $(\varepsilon,\delta)$Freudenthal-Kantor triple 
systems
(abbreviated hereafter 
as to FKTS) [Y-O] offer a simple 
method of constructing Lie algebras (for the case of 
$\delta=+1$) 
and Lie superalgebras 
(for the case of $\delta=-1$) 
with $5$-graded structure, 
it may be of some interest 
to study its symmetry group in 
this note. 
In order to facilitate the discussion,
let 
us briefly sketch its definition.
\par
Let $(V,xyz)$ be a triple 
linear system, 
where $xyz$ 
for 
$x,y,z\in V$
 is a tri-linear product 
in a vector space $V$ over a field $F$.
We introduce two linear mappings $L$ 
and 
$K:V \otimes V\rightarrow {\rm End}\ V$ by
$$
L(x,y)z=xyz,\ K(x,y)z=xzy-\delta yzx
\eqno(1.1)$$
for
$\delta=+1\ {\rm or}\ -1$.
If they satisfy
$$
\begin{array}{l}
[L(u,v),L(x,y)]=L(L(u,v)x,y)+\varepsilon L(x,L(v,u)y),
\hskip 57mm
(1.2)\\
K(K(u,v)x,y)=
L(y,x)K(u,v)-\varepsilon K(u,v)L(x,y)
\hskip 59mm
(1.3)
\end{array}
$$
for any $u,v,x,y\in V$ 
and $\varepsilon=\pm 1,$
we call the triple system to be
$(\varepsilon,\delta)$ FKTS.
\par
One consequence of Eqs.(1.2) and (1.3) is the validity 
of the following important identity
(see [Y-O.] Eqs.(2.9) and (2.10))
$$
\begin{array}{ll}
K(u,v)K(x,y)&
=\varepsilon \delta L(K(u,v)x,y)-
\varepsilon L(K(u,v)y,x)
\hskip 51mm
(1.4)\\
&=L(v,K(x,y)u)-\delta L(u,K(x,y)v).
\hskip 55mm
(1.5)
\end{array}
$$
\par
We can then construct a Lie algebra for
$\delta=+1$
and a Lie superalgebra
for
$\delta=-1$
as follows:
\par
Let $W$ be a space of $2\times 1$
matrix over $V$
$$
W=
\left(\begin{array}{c}
V\\
V
\end{array}
\right)
$$
and
define a tri-linear product:
\par
$W\otimes W\otimes W\rightarrow W$
by
$$
[\left(\begin{array}{c}
x_{1}\\y_{1}
\end{array}
\right),
\left(\begin{array}{c}
x_{2}\\y_{2}
\end{array}
\right),
\left(\begin{array}{c}
x_{3}\\y_{3}
\end{array}
\right)]$$
$$=
\left(\begin{array}{cc}
L(x_{1},y_{2})-\delta L(x_{2},y_{1})&
\delta K(x_{1},x_{2})\\
-\varepsilon K(y_{1},y_{2})&
\varepsilon L(y_{2},x_{1})-\varepsilon\delta L(y_{1},x_{2})
\end{array}
\right)
\left(\begin{array}{c}
x_{3}\\y_{3}
\end{array}
\right).\eqno(1.6)
$$
\par
Then,
it defines a Lie triple system for
$\delta=+1$
and an anti-Lie triple
system
for
$\delta=-1.$
We then note
$$
\hat{L_{{\bar 0}}}=
{\rm span}
\{
\left(
\begin{array}{cc}
L(x,y)&\delta K(z,w)\\
-\varepsilon K(u,v)&\varepsilon L(y,x)
\end{array}
\right)|
x,y,z,w,u,v\in V\}
\eqno(1.7)$$
is a Lie subalgebra of 
${\rm Mat}_{2}({\rm End} (V))^{-},$
where 
$B^{-}$
for an associative algebra $B$
implies a
 Lie algebra with bracket;
$[x,y]=xy-yx.$
We note also then
$$
\hat{D }=
\left(
\begin{array}{cc}
L(x,y),& \delta K(z,w)\\
-\varepsilon K(u,v),& \varepsilon L(y,x)
\end{array}
\right)
\in L(W,W)\eqno(1.8)
$$
is a derivation of the triple
system.
Setting 
$$
L_{\bar{1}}=
{\rm span}
\{
X=
\left(
\begin{array}{c}
x\\y
\end{array}
\right)|
x,y\in V\}
(=W),
\eqno(1.9)
$$
then
$L$ defined by
$$
L=L_{\bar{0}} \oplus
L_{\bar{1}}\eqno(1.10)
$$
gives a Lie algebra for
$\delta=+1,$
and a Lie superalgebra for
$\delta=-1,$
where
$$
L_{\bar{0}}=
\{D|D\ 
{\rm is\ a\ derivation\ of}\ L\},\eqno(1.11)
$$
i.e.,
$D$
satisfies
$$
D[X_{1},X_{2},X_{3}]=
[DX_{1},X_{2},X_{3}]+
[X_{1},DX_{2},X_{3}]+
[X_{1},X_{2},DX_{3}]
\eqno(1.12a)
$$
and hence induces also
$$
[D,[X,Y]]=
[DX,Y]+
[X,DY],\eqno(1.12b)$$
if we define the bracket by
$$
[D_{1}\oplus X_{1},D_{2}\oplus X_{2}]=
([D_{1},D_{2}]+
L(X_{1},X_{2}))\oplus
(D_{1}X_{2}-D_{2}X_{1}).\eqno(1.13)
$$
where
$$
[D_{1},D_{2}]=
D_{1}D_{2}-D_{2}D_{1}$$
and
$$
L(X_{1},X_{2})=
[X_{1},X_{2}]=$$
$$
[\left(
\begin{array}{c}
x_{1}\\
y_{1}
\end{array}
\right),
\left(
\begin{array}{c}
x_{2}\\y_{2}
\end{array}
\right)]=
\left(
\begin{array}{cc}
L(x_{1},y_{2})-\delta L(x_{2},y_{1})&
\delta K(x_{1},x_{2})\\
-\varepsilon K(y_{1},y_{2})&
\varepsilon  L(y_{2},x_{1})-\varepsilon \delta L(y_{1},x_{2})
\end{array}
\right).\eqno(1.14)
$$
\par
Note that the endomorphism
$L(X,Y)$
is then an inner derivation of
the triple system.
\par
Since
$L_{\bar 0}\supset \hat{L}_{\bar{0}},$
we will mainly discuss
a subsystem
$\hat{L}$
of 
$L$, given by
$$
\hat{L}=
\hat{L}_{{\bar 0}}\oplus
L_{{\bar 1}}=
L(W,W)\oplus W\eqno(1.15)
$$
rather than the larger
$L$
except for in section 2.
Then,
$\hat{L}$
is $5$-graded
$$
\hat{L}=
L_{-2}\oplus L_{-1}\oplus
L_{0}\oplus
L_{1}\oplus L_{2}\eqno(1.16)
$$
where
$$
\begin{array}{l}
L_{-2}={\rm span}\{
\left(
\begin{array}{cc}
0&0\\
-\varepsilon K(x,y)&0
\end{array}
\right)|
x,y\in V\}\hskip 68mm
(1.17a)\\
L_{-1}={\rm span}\{
\left(
\begin{array}{c}
0\\
x
\end{array}
\right)|
x\in V\}
\hskip 96mm
(1.17b)\\
L_{0}={\rm span}
\{
\left(
\begin{array}{cc}
L(x,y)&0\\
0&\varepsilon L(y,x)
\end{array}
\right)|
x,y\in V\}
\hskip 65mm
(1.17c)\\
L_{1}={\rm span}
\{
\left(
\begin{array}{c}
x\\0
\end{array}\right)|
x\in V\}
\hskip 98mm
(1.17d)\\
L_{2}={\rm span}\{
\left(
\begin{array}{cc}
0&\delta K(x,y)\\
0&0
\end{array}\right)|
x,y\in V\}.
\hskip 74.5mm
(1.17e)
\end{array}
$$
\par
Here, we utilized the following Proposition for some of its proof.
\par
\vskip 3mm
{\bf Proposition 1.1}
([K-O.],
[K-M-O.])
\par
\vskip 3mm
\it
Let $(V, \ (xyz))$ be a
$(\varepsilon,\delta)$
-Freudenthal-Kantor triple system
with an endomorphism
$P$
such that
$P^{2}=-\varepsilon\delta {\rm Id}$
and
$P(xyz)=
(PxPyPz).$
Then,
$(V,[xyz])$
is a Lie triple system
(for $\delta=1$)
and anti-Lie triple system
(for $\delta=-1$)
with respect to the product
$$
[xyz]=
(xPyz)-\delta(yPxz)+
\delta(xPzy)-
(yPzx).
$$
\rm
In passing, we note that the standard 
$\hat{L}=\Sigma_{i=-2}^{2}\oplus L_{i}$
is a result of 
Proposition
$1.1$ immediately
with
$$
P=
\left(
\begin{array}{cc}
0,&\delta\\
-\varepsilon ,&0
\end{array}
\right)
\ \ and\ 
x\rightarrow X \ etc.
$$

\par
Next,
we introduce
$\theta,\sigma(\lambda)\in {\rm End}(\hat{L})$
for any
$\lambda\in F
(\lambda\not= 0),
$
being non-zero constant by
$$
\theta
\left(\begin{array}{c}
x\\y
\end{array}
\right)=
\left(\begin{array}{c}
-\varepsilon y\\
\delta x
\end{array}
\right)\eqno(1.18a)
$$
$$
\sigma(\lambda)
\left(
\begin{array}{c}
x\\ y
\end{array}
\right)
=
\left(\begin{array}{c}
\lambda x\\
{\frac{1}{\lambda}}
y
\end{array}
\right)\eqno(1.18b)
$$
in
$W=L_{\bar{1}}=L_{-1}\oplus L_{1}.$
We may easily verify that
they are automorphism of
$
[W,W,W],$
i.e,
we have for example
$$
\theta([X,Y,Z])=
[\theta X,\theta Y,\theta Z]
$$
for
$X,Y,Z\in W.$
We then extend their actions to the whole of
$\hat{L}$
in a natural way to show that they
will define automorphism
of
$\hat{L}.$
They moreover satisfy
$$
\begin{array}{ll}
(i)&
 \sigma(1)=Id,\theta^{4}=Id\hskip 101mm(1.19a)\\
 &
{\rm where\ {\it Id}\ is\ the\ 
identity\ mapping}\\
(ii)
&
\theta^{2}=-\varepsilon \delta Id
\ {\rm for}
L_{\bar{1}}\ but\ 
\theta^{2}=Id\ {\rm for}\ 
L_{\bar{0}}
\hskip 69mm(1.19b)\\
(iii)&
\sigma(\mu)\sigma(\nu)=\sigma(\mu\nu)\ {\rm for}\ 
\mu,\nu\in F,\ \mu\nu\not= 0\hskip 64mm(1.19c)\\
(iv)&
\sigma(\lambda)\theta\sigma(\lambda)=\theta \ {\rm for\ 
any}
\ \lambda\in F,\lambda\not=0.\hskip 69.5mm(1.19d)
\end{array}
$$
We call the group generated by
$\sigma(\lambda)$ and $\theta$
satisfying these
conditions simply as
$D(\varepsilon,\delta)$
due to a lack of better
terminology.
If the field
$F$ contains
$\omega\in F$ satisfying
$\omega^{3}=1$
but
$\omega\not=1,$
then a finite sub-group of
$D(\varepsilon,\delta)$
generated by 
$\theta$ and
$\sigma(\omega)$
defines
$Dic_{3}$
group for
$\varepsilon=\delta$
but
$S_{3}$
for
$\varepsilon=-\delta,$
as we have noted already
([E-O,4]).
\par
Conversly any
$5$-graded Lie algebra
(or Lie superalgebra)
with such automorphism
$\theta$ and
$\sigma(\lambda)$ satisfying
Eqs.(1.19)
lead
essentially
to a
$(\varepsilon,\delta)$FKTS
in $L_{1}$
with a triple product defined
by
$\{x,y,z\}=[[x,\theta y],z]$
for
$x,y,z\in L_{1}$
(see [E-K-O]).
\par
We note that the
corresponding  local symmetry of
$D(\varepsilon,\delta)$
yields a derivation of
$\hat{L},$
given by
$$
h=\left(
\begin{array}{cc}
1&0\\
0&-1
\end{array}
\right)\eqno(1.20)
$$
which satisfies
$$
h[X,Y,Z]=
[hX,Y,Z]+[X,hY,Z]+
[X,Y,hZ]\eqno(1.21a)$$
as well as
$$
[h,[X,Y]]=
[hX,Y]+
[X,hY]\eqno(1.21b)$$
for
$X,Y,Z\in W.$
\par
We can find a larger
automorphism group of
$\hat{L},$
if we impose some additional conditions.
First suppose that
$K(x,y)$
is now expresed as
$$
K(x,y)=
\varepsilon\delta L(y,x)-
\varepsilon L(x,y)\eqno(1.22)$$
for any
$x,y \in V.$
We call then the triple system
 to be a special
 $(\varepsilon,\delta)$FKTS
 ([E-O,4]).
 As we will see at the end of section 2,
 this triple system is intimately related
 to zero
 Nijenhuis tensor condition.
 Moreover for the case of special
 $(\varepsilon,\varepsilon)$FKTS
 (i.e. $\varepsilon=\delta$),
 the automorphism group of
 $\hat{L}$
 turns out to be a larger
 $SL(2,F) (=Sp(2,F))$
 group which contains
 $D(\varepsilon,\varepsilon)$
  as its subgroup.
   In this case,
 the triple system
 $[W,W,W]$
 becomes invariant under
 $$
 \left(\begin{array}{c}
 x\\y
 \end{array}
 \right)
 \rightarrow
 U
 \left(\begin{array}{c}
 x
\\y
\end{array}
\right)\eqno(1.23a)$$
for any
$2\times 2\ SL(2,F)\ {\rm matrix}\ U,\ {\rm i.e.}$
$$
U=\left(\begin{array}{cc}
\alpha&\beta\\
\mu& \nu
\end{array}
\right),
{\rm Det}\ U=\alpha\nu-\beta\mu=1,\eqno(1.23b)$$
as we will show in the next section
(Sction 2).
Also,
the associated
Lie algebras or superalgebras are
$BC_{1}$-graded algebra of
type $C_{1}.$
\par
Finally (Section 3),
we consider a ternary 
system
$(V,xy,xyz)$
where
$xy$
and
$xyz$
are binary and ternary
products,
respectively,
in the vector space
$V$.
Suppose that they satisfy
\par
(1)\ the triple system
$(V,xyz)$ is a $(-1,1)$FKTS.
\par
(2)\ The binary algebra
$(V,xy)$
is unital and involution 
$({\overline{xy}}={\bar y}{\bar x})$
with
 the
involutive map
$x\rightarrow {\bar x},{\bar{\bar x}}=x.$
\par
(3)\ The triple product $xyz$ is expressed in terms of the
bi-linear products by
$$
xyz=(z{\bar y})x-
(z{\bar x})y+(x{\bar y})z.
\eqno (1.24)$$
\par
We may call the ternary system
$(V,xy,xyz)$
to be Allison-ternary algebra
or simply $A$-ternary algebra,
since $A=(V,xy)$
is then the structurable algebra
([A],[A-F]).
\par
This case is of great interest,
first because structurable algebras exhbit a triality 
relation
([A-F]),
and second because we can construct another type
of Lie algebras
independently of the
standard construction of
$(-1,1)$ FKTS,
which is $S_{4}$-invariant and of
$BC_{1}$ graded Lie algebra of type $B_{1}.$
The relationship between the Lie algebra constructed in
the new way and that given as in Eq.(1.17)
is by no means transparent.
Note that the group
$D(-1,1)$ contains $S_{3}$ but not $S_{4}$
symmetry.
In section 3,
we will show that if the field $F$ contains the square root
$\sqrt{-1}$ of $-1$,
then Eqs.(1.17)
can be prolonged to yield the Lie algebra for 
the structurable algebra.
\vskip 3mm
{\bf 2.\ Symmetry Group of Lie Algebras associated with
$(\varepsilon ,\delta)$FKTS}
\vskip 3mm
Although the invariance of the Lie algebra
or superalgebra $\hat{L}$
under $\theta$
and $\sigma(\lambda)$
given by Eqs.(1.18)
has been already noted in
[E-O,4]
let us recapitulate its proof briefly as follows:
For $\theta$ given by
Eq.(1.18a),
it is easy to verify the validity of
$$
\theta([X_{1},X_{2},X_{3}])=
[\theta X_{1},\theta X_{2},\theta X_{3}]$$
for
$X_{i}=
\left(
\begin{array}{c}
x_{j}\\y_{j}
\end{array}
\right) \in W\ (j=1,2,3).$
Since
$\hat{L}$ is $5$-graded,
it is also invariant under
$$
\sigma_{n}(\lambda):
Z_{n}\rightarrow \lambda^{n}Z_{n},\ 
(n=0,\pm 1,\pm 2)$$
for any
$Z_{n}\in L_{n}$
given in Eqs.(1.17).
This implies the validity
of Eq.(1.18b).
Then,
Eqs.(1.19)
can be readily verified.
Thus, a generic Lie algebra or superalgebra
$\hat{L}$
associated with
$(\varepsilon,\delta)$FKTS has the
symmetry group
$D(\varepsilon,\delta)$ generated
by $\theta$ and
$\sigma(\lambda)$
satisfing Eqs.(1.19).
However,
for some special
$(\varepsilon,\delta)$FKTS,
the invariance group for
$\hat{L}$ can be larger
as follows.
\par
Let us consider the case of special
$(\varepsilon,\varepsilon)$FKTS
([E-O,4]),
where
$K(x,y)$
is expressed as
$$
K(x,y)=L(y,x)-\varepsilon L(x,y)\eqno(2.1)$$
in terms of $L(x,y)'s.$
We can show that
$\sigma\in {\rm End}\ \hat{L}$
defined by
$$
\sigma
\left(
\begin{array}{c}
x\\y
\end{array}
\right):=
U
\left(
\begin{array}{c}
x\\y
\end{array}
\right)
=
\left(
\begin{array}{cc}
\alpha&\beta\\
\mu&\nu
\end{array}
\right)
\left(
\begin{array}{c}
x\\y
\end{array}
\right)=
\left(
\begin{array}{c}
\alpha x+\beta y\\
\mu x+\nu y
\end{array}
\right)
\eqno(2.2)$$
gives an automorphism of the Lie algebra or superalgebra
$\hat{L}$,
provided that we have
$$
Det U=\alpha \nu-\beta\mu=1.\eqno(2.3)
$$
\par
We further
 define the action of
$\sigma$ on $L(W,W)$
by
$$
\sigma
\left([
\left(
\begin{array}{c}
x_{1}\\y_{1}
\end{array}
\right),
\left(
\begin{array}{c}
x_{2}\\y_{2}
\end{array}
\right)
]
\right)=
[\sigma
\left(
\begin{array}{c}
x_{1}\\y_{1}
\end{array}
\right),
\sigma
\left(
\begin{array}{c}
x_{2}\\y_{2}
\end{array}
\right)],\eqno(2.4)
$$
and  will prove   
the following.
\par
\vskip 3mm
{\bf Proposition 2.1}
\par
\vskip 3mm
\it
Under the assumption as in above,
we have
$$
\sigma([
\left(\begin{array}{c}
x_{1}\\y_{1}
\end{array}
\right),
\left(\begin{array}{c}
x_{2}\\y_{2}
\end{array}
\right)])=
[\sigma
\left(\begin{array}{c}
x_{1}\\y_{1}
\end{array}
\right),
\sigma\left(\begin{array}{c}
x_{2}\\y_{2}
\end{array}
\right)]=
U[
\left(\begin{array}{c}
x_{1}\\y_{1}
\end{array}
\right),
\left(\begin{array}{c}
x_{2}\\y_{2}
\end{array}
\right)]
U^{-1}.\eqno(2.5)$$
\par
\rm
\vskip 3mm
{\bf Proof}
\par
\vskip 3mm
First we note
$$
[
\left(\begin{array}{c}
x_{1}\\y_{1}\end{array}
\right),
\left(\begin{array}{c}
x_{2}\\y_{2}
\end{array}
\right)]=
\left(
\begin{array}{cc}
L(x_{1},y_{2})-\varepsilon L(x_{2},y_{1}),&
\varepsilon K(x_{1},x_{2})\\
-\varepsilon K(y_{1},y_{2}),&
\varepsilon L(y_{2},x_{1})-L(y_{1},x_{2})
\end{array}
\right)
\eqno(2.6)
$$
since
$\varepsilon=\delta,$
so that the
right side
Eq.(2.4)
is calculated to be 
$$
[\sigma
\left(\begin{array}{c}
x_{1}\\
y_{1}
\end{array}
\right),
\sigma
\left(
\begin{array}{c}
x_{2}\\
y_{2}
\end{array}
\right)]
=
[
\left(
\begin{array}{c}
\alpha x_{1}+\beta y_{1}\\
\mu x_{1}+\nu y_{1}
\end{array}
\right),
\left(
\begin{array}{c}
\alpha x_{2}+\beta y_{2}\\
\mu x_{2}+\nu y_{2}
\end{array}
\right)]=
\left(\begin{array}{cc}
A&B\\
C&D
\end{array}
\right)
\eqno(2.7)
$$
where we have set
$$
\begin{array}{ll}
A&
=L(\alpha x_{1}+\beta y_{1},\mu x_{2}+\nu y_{2})-
\varepsilon L(\alpha x_{2}+\beta y_{2},
\mu x_{1}+\nu y_{1})\\
&=
\alpha\mu\{L(x_{1},x_{2})-\varepsilon L(x_{2},x_{1}\}+
\beta \nu\{L(y_{1},y_{2})-
\varepsilon L(y_{2},y_{1})\\
&+
\{\alpha \nu L(x_{1},y_{2})-\varepsilon
\beta\mu
L(y_{2},x_{1})\}+
\{\beta \nu L(y_{1},x_{2})
-\varepsilon \alpha \nu L(x_{2},y_{1})\}\\
&=
-\alpha\mu\varepsilon K(x_{1},x_{2})-
\beta \nu
\varepsilon
K(y_{1},y_{2})\\
&+
\{\alpha \nu L(x_{1},y_{2})
-\varepsilon\beta \mu L(y_{2},x_{1})\}+
\{\beta \nu L(y_{1},x_{2})-
\varepsilon\alpha \nu 
L(x_{2},y_{1})\}\hskip 28.5mm
(2.8a)\\
D&
=\varepsilon L(\mu x_{2}+\nu y_{2},
\alpha x_{1}+\beta y_{1})-
L(\mu x_{1}+\nu y_{1},
\alpha x_{2}+\beta y_{2})\\
&=
\mu \alpha(\varepsilon L(x_{2},x_{1})-
L(x_{1},x_{2}))+
\nu\beta(\varepsilon L(y_{2},y_{1})-L(y_{1},y_{2}))\\
&+
\{\nu\alpha\varepsilon
L(y_{2},x_{1})
-\mu\beta L(x_{1},y_{2})\}+
\{\mu\beta\varepsilon L(x_{2},y_{1})
-\nu\alpha L(y_{1},x_{2})\}\hskip 28mm
(2.8b)\\
&=
\mu\alpha\varepsilon K
(x_{1},x_{2})+
\nu\beta \varepsilon K(y_{1},y_{2})+
\nu\alpha\varepsilon L(y_{2},x_{1})
\\
&-
\mu\beta L(x_{1},y_{2})+
\mu\beta\varepsilon L
(x_{2},y_{1})
-\nu\alpha L(y_{1},x_{2})\\
B&
=\varepsilon\alpha^{2}K(x_{1},x_{2})+
\varepsilon\beta^{2}K(y_{1},y_{2})+
\varepsilon\alpha\beta K(x_{1},y_{2})+
\varepsilon\alpha\beta K(y_{1},x_{2}),
\hskip 26.5mm
(2.8c)\\
C&
=-\varepsilon\mu^{2}K(x_{1},x_{2})
-\varepsilon \nu^{2}K(y_{1},y_{2})
-\varepsilon\mu\nu K(x_{1},y_{2})
-\varepsilon\mu \nu K(y_{1},x_{2})
\hskip 26mm
(2.8d)
\end{array}
$$
\par
Here,
we used Eq.(2.1)
to simplify the last lines
in Eqs.(2.8a) and (2.8b).
\par
Then,
we find further
$$
\begin{array}{l}
\alpha A+\mu B=\alpha(L(x_{1},y_{2})-\varepsilon L(x_{2},y_{1}))-
\beta\varepsilon K(y_{1},y_{2})\\
\beta A+
\nu B=
\beta(\varepsilon L(y_{2},x_{1})-
L(y_{1},x_{2})+
\varepsilon\alpha K(x_{1},x_{2})\\
\alpha C+\mu D=
\mu(L(x_{1},y_{2})-
\varepsilon L(x_{2},y_{1}))-
\varepsilon \nu K(y_{1},y_{2})\\
\beta C+\nu D=
\nu(\varepsilon L(y_{2},x_{1})-
L(y_{1},x_{2}))+
\varepsilon \mu K(x_{1},x_{2})
\end{array}
$$
in view of $\alpha\nu-\beta\mu=1,$
so that we have 
$$
[\sigma
\left(
\begin{array}{c}
x_{1}\\y_{1}
\end{array}
\right),
\sigma
\left(\begin{array}{c}
x_{2}\\y_{2}
\end{array}
\right)
]
U=
\left(
\begin{array}{cc}
A&B\\
C&D
\end{array}
\right)
\left(\begin{array}{cc}
\alpha&\beta\\
\mu&\nu
\end{array}
\right)=
\left(
\begin{array}{cc}
\alpha A+\mu B &
\beta A+\nu B\\
\alpha C+\mu D &\beta C+\nu D
\end{array}
\right)
$$
which is rewritten further as
$$
[\sigma
\left(\begin{array}{c}
x_{1}\\y_{1}
\end{array}
\right),
\sigma
\left(
\begin{array}{c}
x_{2}\\y_{2}
\end{array}
\right)]U
=
\left(\begin{array}{cc}
\alpha&\beta\\
\mu&\nu
\end{array}
\right)
\left(
\begin{array}{cc}
L(x_{1},y_{2})-\varepsilon L(x_{2},y_{1})&
\varepsilon K(x_{1},x_{2})\\
-\varepsilon K(y_{1},y_{2})&
\varepsilon L(y_{2},x_{1})-L(y_{1},x_{1})
\end{array}
\right)
$$
$$
=U
[
\left(
\begin{array}{c}
x_{1}\\y_{1}
\end{array}
\right),
\left(
\begin{array}{c}
x_{2}\\y_{2}
\end{array}
\right)]
.
$$

This proves Eq.(2.5), and completes the proof.//
\par
Then this yields also

$$
\sigma([
X_{1},X_{2},X_{3}])=
[\sigma X_{1},\sigma X_{2},\sigma X_{3}]$$
for
$X_{j}=\left(
\begin{array}{c}
x_{j}\\
y_{j}
\end{array}
\right)
\in W\quad
(j=1,2,3),$
since we calculate
$$
\sigma [X_{1},X_{2},X_{3}]=U[X_{1},X_{2},X_{3}]\ \ \ (since \ 
[X_{1},X_{2},X_{3}] \in W\ )
$$
$$
=U[X_{1},X_{2}]X_{3}\ \ \ (\ by \ Eqs. \ (1.6)\ and \ (1.14)\ )
$$
$$
=
U[X_{1},X_{2}]U^{-1} U X_{3}\ \ \ (\ since \ [X_{1},X_{2}] \in 
L(W,W) \ is \ a \ 2 \times 2\ matrix\ )
$$
$$
=[\sigma X_{1},\sigma X_{2}]\sigma X_{3}\ \ \ (\ by \ Eqs.\ (2.2)\ and\ (2.5)\ ).
$$

\par
In conclution,
the Lie algebra or superalgebra
$\hat{L}$
constructed from
any special
$(\varepsilon,\varepsilon)$FKTS
admits
$SL(2)(=Sp(2))$
as its automorphism group.
Note that
$SL(2)$ contains the group
$D(\varepsilon,\varepsilon)$
by
$$
D(\varepsilon,\varepsilon)
\rightarrow SL(2)$$
$$
\theta\rightarrow
\left(
\begin{array}{cc}
0 &-\varepsilon\\
\varepsilon& 0
\end{array}
\right)
$$
$$
\sigma(\lambda)
\rightarrow
\left(
\begin{array}{cc}
\lambda &0\\
0 & {\frac{1}{\lambda}}
\end{array}
\right).$$
\par
We then define an operator
$\hat{\sigma}$
on
$\hat{L}$
as follows:
$$
\left(
\begin{array}{cc}
L(a,b)&\varepsilon K(c,d)\\
-\varepsilon K(e,f)&
\varepsilon L(b,a)
\end{array}
\right)
\oplus
\left(
\begin{array}{c}
x\\y
\end{array}
\right)
\rightarrow
U
\left(
\begin{array}{cc}
L(a,b)&\varepsilon K(c,d)\\
-\varepsilon K(e,f)&
\varepsilon L(b,a)
\end{array}
\right)
U^{-1}
\oplus
U
\left(\begin{array}{c}
x\\y
\end{array}
\right)$$
where
$$
\sigma:=
U=
\left(\begin{array}{cc}
\alpha&\beta\\
\mu&\nu
\end{array}
\right)\ 
{\rm and}\ 
det\ U=1.
$$
Then
$\hat{\sigma}$
is an
automorphism of
$\hat{L}$
induced from
the triple system.
\par
Moreover as the local version of the global
$SL(2)$ symmetry,
we can prove also the
following.
\par
\vskip 3mm
{\bf Proposition 2.2}
\par
\vskip 3mm
\it
Under the assumption as in above,
let
$$
h=\left(
\begin{array}{cc}
1&0\\
0&-1
\end{array}
\right),
f=
\left(
\begin{array}{cc}
0&1\\
0&0
\end{array}
\right),
g=
\left(
\begin{array}{cc}
0&0\\
1&0
\end{array}
\right)
\eqno(2.9)
$$
which forms a $sl(2)$ 
Lie algebra by
the standard commutation relations.
We also
define their actions to
$
X=
\left(
\begin{array}{c}
x\\y
\end{array}
\right)
\in W$
by
$$
[A,X]=A
\left(
\begin{array}{c}
x\\y
\end{array}
\right)
$$
for
$A=h,f$
or
$g$ and
similarly define
$[A,M]=
AM-MA$
for 
$M\in L(W,W)$.
\par
Then,
$h,f$ and
$g$
are derivations of
$\hat{L}=
L(W,W)\oplus W$
for
special
$(\varepsilon,\varepsilon)$FKTS 
and vice versa.
\par 
\vskip 3mm
{\bf Proof.}
\par
\vskip 3mm
\rm
First,
$h$ is actually
a derivation of
$\hat{L}$ for
any $(\varepsilon,\delta)$FKTS
as we have already noted.
Also,
the fact that
$g$ is a derivation of
$\hat{L}$ for special
$(\varepsilon,\varepsilon)$FKTS
has been  proven in
[K-O]
(see Theorem 5.2).
Here we will show similarly the validity of
$$
f[X_{1},X_{2},X_{3}]=
[fX_{1},X_{2},X_{3}]+
[X_{1},fX_{2},X_{3}]+
[X_{1},X_{2},fX_{3}]\eqno(2.10)$$
which prove these being  derivations of $W$.
Noting
$$
f
\left(
\begin{array}{c}
x\\y
\end{array}
\right)=
\left(
\begin{array}{cc}
0&1\\
0&0
\end{array}
\right)
\left(\begin{array}{c}
x\\y
\end{array}
\right)=
\left(
\begin{array}{c}
y\\0
\end{array}
\right),
$$
\par
we then calculate
$$
\begin{array}{l}
f [
\left(
\begin{array}{c}
x_{1}\\y_{1}
\end{array}
\right),
\left(
\begin{array}{c}
x_{2}\\y_{2}
\end{array}
\right),
\left(
\begin{array}{c}
x_{3}\\y_{3}
\end{array}
\right)]\\
=
\left(
\begin{array}{cc}
0&1\\
0&0
\end{array}
\right)
\left(
\begin{array}{cc}
L(x_{1},y_{2})-\varepsilon
L(x_{2},y_{1}),&
\varepsilon K(x_{1},x_{2})\\
-\varepsilon K(y_{1},y_{2}),&
\varepsilon L(y_{2},x_{1})-
L(y_{1},x_{2})
\end{array}
\right)
\left(
\begin{array}{c}
x_{3}\\
y_{3}
\end{array}
\right)\\
=
\left(
\begin{array}{cc}
-\varepsilon K(y_{1},y_{2}),&
\varepsilon L(y_{2},x_{1})-
L(y_{1},x_{1})\\
0,&0
\end{array}
\right)
\left(
\begin{array}{c}
x_{3}\\
y_{3}
\end{array}
\right)\\
=
\left(
\begin{array}{c}
-\varepsilon K(y_{1},y_{2})x_{3}+
(\varepsilon L(y_{2},x_{1})-
L(y_{1},x_{2}))y_{3}\\
0
\end{array}
\right).
\end{array}
$$
\par
But
$$
[f
\left(
\begin{array}{c}
x_{1}\\y_{1}
\end{array}
\right),
\left(
\begin{array}{c}
x_{2}\\
y_{2}
\end{array}
\right),
\left(
\begin{array}{c}
x_{3}\\y_{3}
\end{array}
\right)]=
[
\left(
\begin{array}{c}
y_{1}\\0
\end{array}
\right),
\left(
\begin{array}{c}
x_{2}\\y_{2}
\end{array}
\right),
\left(
\begin{array}{c}
x_{3}\\y_{3}
\end{array}
\right)]
$$
$$
=
\left(
\begin{array}{cc}
L(y_{1},y_{2})&
\varepsilon K(y_{1},x_{2})\\
0&\varepsilon L(y_{2},y_{1})
\end{array}
\right)
\left(
\begin{array}{c}
x_{3}\\
y_{3}
\end{array}
\right)
=
\left(
\begin{array}{c}
L(y_{1},y_{2})x_{3}+\varepsilon K(y_{1},x_{2})y_{3}\\
\varepsilon L(y_{2},y_{1})y_{3}
\end{array}
\right),
$$
$$
[
\left(
\begin{array}{c}
x_{1}\\y_{1}
\end{array}
\right),
f
\left(
\begin{array}{c}
x_{2}\\
y_{2}
\end{array}
\right),
\left(
\begin{array}{c}
x_{3}\\
y_{3}
\end{array}
\right)]
=
[
\left(
\begin{array}{c}
x_{1}\\y_{1}
\end{array}
\right),
\left(
\begin{array}{c}
y_{2}\\0
\end{array}
\right),
\left(
\begin{array}{c}
x_{3}\\y_{3}
\end{array}
\right)]
$$
$$
=
\left(
\begin{array}{cc}
-\varepsilon L(y_{2},y_{1}),&
\varepsilon K(x_{1},y_{2})\\
0,&
-L(y_{1},y_{2})
\end{array}
\right)
\left(
\begin{array}{c}
x_{3}\\y_{3}
\end{array}
\right)
=
\left(
\begin{array}{c}
-\varepsilon L(y_{2},y_{1})x_{3}+
\varepsilon K(x_{1},y_{2})y_{3}\\
-L(y_{1},y_{2})y_{3}
\end{array}
\right),
$$
$$
[
\left(
\begin{array}{c}
x_{1}\\y_{1}
\end{array}
\right),
\left(
\begin{array}{c}
x_{2}\\
y_{2}
\end{array}
\right),
f
\left(
\begin{array}{c}
x_{3}\\y_{3}
\end{array}
\right)]
=
[
\left(
\begin{array}{c}
x_{1}\\y_{1}
\end{array}
\right),
\left(
\begin{array}{c}
x_{2}\\y_{2}
\end{array}
\right),
\left(
\begin{array}{c}
y_{3}\\0
\end{array}
\right)]
$$
$$
=
\left(
\begin{array}{cc}
L(x_{1},y_{2})-\varepsilon L(x_{2},y_{1}),&
\varepsilon K(x_{1},x_{2})\\
-\varepsilon K(y_{1},y_{2}),&
\varepsilon L(y_{2},x_{1})-L(y_{1},x_{2})
\end{array}
\right)
\left(
\begin{array}{c}
y_{3}\\0
\end{array}
\right)
$$
$$
=
\left(
\begin{array}{c}
(L(x_{1},y_{2})-\varepsilon L(x_{2},y_{1}))y_{3}\\
-\varepsilon K(y_{1},y_{2})y_{3}
\end{array}
\right).
$$
\par
Hence,
setting
$$
\begin{array}{l}
[f
\left(
\begin{array}{c}
x_{1}\\y_{1}
\end{array}
\right),
\left(
\begin{array}{c}
x_{2}\\y_{2}
\end{array}
\right),
\left(
\begin{array}{c}
x_{3}\\y_{3}
\end{array}
\right)]
+
[
\left(
\begin{array}{c}
x_{1}\\y_{1}
\end{array}
\right),
f
\left(
\begin{array}{c}
x_{2}\\y_{2}
\end{array}
\right),
\left(
\begin{array}{c}
x_{3}\\
y_{3}
\end{array}
\right)]\\
+
[
\left(
\begin{array}{c}
x_{1}\\y_{1}
\end{array}
\right),
\left(
\begin{array}{c}
x_{2}\\y_{2}
\end{array}
\right),
f
\left(
\begin{array}{c}
x_{3}\\y_{3}
\end{array}
\right)]
=
\left(
\begin{array}{c}
z\\w
\end{array}
\right),
\end{array}
$$
we calculate
$$
\begin{array}{ll}
z
&=
L(y_{1},y_{2})x_{3}+
\varepsilon K(y_{1},x_{2})y_{3}-
\varepsilon L(y_{2},y_{1})x_{3}+
\varepsilon K(x_{1},y_{2})y_{3}\\
&+
(L(x_{1},y_{2})-
\varepsilon L(x_{2},y_{1}))y_{3}\\
&=
\{L(y_{1},y_{2})-
\varepsilon L(y_{2},y_{1})\}x_{3}+
\{\varepsilon K(y_{1},x_{2})+
\varepsilon K(x_{1},y_{2})+
L(x_{1},y_{2})-
\varepsilon L(x_{2},y_{1})\}y_{3}\\
&=
K(y_{2},y_{1})x_{3}+
\{\varepsilon(L(x_{2},y_{1})-
\varepsilon L(y_{1},x_{2}))\}y_{3}\\
&+\{
\varepsilon(L(y_{2},x_{1})-
\varepsilon L(x_{1},y_{2}))+
L(x_{1},y_{2})-
\varepsilon L(x_{2},y_{1})\}y_{3}\\
&=
-\varepsilon K(y_{1},y_{2})x_{3}+
\{-L(y_{1},x_{2})+
\varepsilon L(y_{2},x_{1})\}y_{3}
\end{array}
$$
and
$$\begin{array}{ll}
w&
=
\varepsilon L(y_{2},y_{1})y_{3}-
L(y_{1},y_{2})y_{3}-
\varepsilon K(y_{1},y_{2})y_{3}\\
&
=\{
\varepsilon L(y_{2},y_{1})-
L(y_{1},y_{2})-
\varepsilon K(y_{1},y_{2})\}
y_{3}=0.
\end{array}
$$
Comparing these with the left hand of
Eq.(2.10),
we obtain
$$
f[X_{1},X_{2},X_{3}]=
[fX_{1},X_{2},X_{3}]+
[X_{1},fX_{2},X_{3}]+
[X_{1},X_{2},fX_{3}],
$$
so that
$f$ is also
a derivation
of the associated Lie algebra
$\hat{L}$.
This completes the proof of Proposition 2.2.//
\par
\vskip 3mm
{\bf Remark.2.3}
\par
\vskip 3mm
Since $sl(2)=(h,f,g)$
are derivations of
$\hat{L}=L(W,W)\oplus W,$
we can add them to enlarge
the Lie algebra or supleralgebra
$\hat{L}$
into
$L=\hat{L}\oplus sl(2)$
by Eq.(1.13).
\par
Then,
$L$
as modules of $sl(2)$
is a direct sum of one,
two,
and three dimensional
modules,
as one can be easily seen
as follows:
\par
(1)\quad
Two-dimensional modules consist of
$$
\left(
\begin{array}{c}
x\\0
\end{array}
\right)
\ {\rm and}\ 
\left(
\begin{array}{c}
0\\x
\end{array}
\right)
$$
\par
(2)\quad
Three-dimensional modules consist of 
$sl(2)$ and also of   
$$
\begin{array}{l}
\left(
\begin{array}{cc}
0&K(x,y)\\
0&0
\end{array}
\right),
\left(
\begin{array}{cc}
0&0\\
K(x,y)&0
\end{array}
\right), \ and,\\
\left(
\begin{array}{cc}
K(x,y)&0\\
0&-K(x,y)
\end{array}
\right)
=
\left(
\begin{array}{cc}
L(y,x)&0\\
0&\varepsilon L(x,y)
\end{array}
\right)
-\varepsilon
\left(
\begin{array}{cc}
L(x,y)&0\\
0&\varepsilon L(y,x)
\end{array}
\right)
\in L(W,W)
\end{array}
$$
\par
(3)\quad
Trivial modules consist of
$$
\begin{array}{l}
\left(
\begin{array}{cc}
L(y,x)+\varepsilon L(x,y)&0\\
0&
L(y,x)+\varepsilon L(x,y)
\end{array}
\right)\\
=
\left(
\begin{array}{cc}
L(y,x)&0\\
0&
\varepsilon L(x,y)
\end{array}
\right)
+
\varepsilon
\left(
\begin{array}{cc}
L(x,y)&0\\
0&\varepsilon L(y,x)
\end{array}
\right)
\in L(W,W)
\end{array}
$$
\par
so that
$L$ is a 
$BC_{1}$-graded Lie
algebra or superalgebra of type
$C_{1}.$
This fact is in essential accord with
results of
Corollaries 3.8 and 4.6 of [E-O,4],
which are
based upon analysis of the $J$-ternary algebra
[A-B-G].
\par
\vskip 3mm
{\bf Remark 2.4}\par
\vskip 3mm
A $(\varepsilon,\varepsilon)$ FKTS is called
unitary (see [K-M-O]),
if
$K(V,V)$
contains an identity map,
i.e.,
there exist
$a_{i},b_{i}\in V$
satisfying
$$
\sum_{i} K(a_{i},b_{i})=Id.
$$
\par
Any unitary
$(\varepsilon,\varepsilon)$ FKTS
is special
([K-M-O],[E-O,4]).
Moreover
$(h,g,f)$
constructed 
above are now
contained in
$L(W,W)$
by replacing
$K(x,y)$ in Remark 2.3 by
$\sum_{i}, K(a_{i},b_{i})=Id.$
Further a
$(\varepsilon,\varepsilon)$ FKTS
is said to be balanced if we have
$K(x,y)=<x|y>Id$
for a non-zero bi-linear form
$<\cdot|\cdot>.$
Then,
any balanced 
$(\varepsilon,\varepsilon)$ FKTS
is unitary and hence speical.
If the field $F$ is algebraically 
closed of zero characteristic,
any simple Lie algebra except for
$sl(2)$ can be 
constructed
standardly from some
balanced
$(1,1)$FKTS
([M],[Ka],[E-K-O])
so that any such simple 
classical Lie algebra is automatically
$BC_{1}$-graded Lie algebra of 
type $C_{1}$.
\par
\vskip 3mm
{\bf Remark 2.5}\par
\vskip 3mm
Let us set
(for any $(\varepsilon,\delta)$FKTS):
$$
J=f-g=
\left(
\begin{array}{cc}
0&1\\
-1&0
\end{array}
\right).\eqno(2.11)
$$
\par
We can then verify the validity of
$$
J[X,Y]J^{-1}=
\varepsilon \delta[JX,JY],\eqno(2.12a)
$$
as well as
$$
J[X,Y,Z]=
\varepsilon\delta
[JX,JY,JZ]
\eqno(2.12b)
$$
for any
$X,Y,Z\in W$.
We next introduce an analogue of Nijenhuis tensor 
in diffrential geometry
([K-N])
by
$$
N(X,Y)=
[JX,JY]-
J[JX,Y]-
J[X,JY]+
J^{2}[X,Y].\eqno(2.13)
$$
Setting
$
X=\left(
\begin{array}{c}
x_{1}\\
y_{1}
\end{array}
\right),$
and
$
Y=
\left(
\begin{array}{cc}
x_{2}\\
y_{2}
\end{array}
\right),
$
we then calculate
$$
N(X,Y)=
\left(
\begin{array}{cc}
-\varepsilon(\Lambda(x_{1},y_{2})+
\Lambda(y_{1},x_{2})),&
\delta(\Lambda(y_{1},y_{2})-
\Lambda(x_{1},x_{2}))\\
\varepsilon(\Lambda(y_{1},y_{2})-
\Lambda(x_{1},x_{2})),&
\delta(\Lambda(x_{1},y_{2})+
\Lambda(y_{1},x_{2}))
\end{array}
\right),\eqno(2.14)
$$
where
$$
\Lambda(x,y)=
K(x,y)+
\varepsilon L(x,y)-
\varepsilon\delta L(x,y).\eqno(2.15)
$$
\par
Therefore,
for any special
$(\varepsilon,\delta)$
FKTS
(which is defined by
$\Lambda(x,y)=0,$
see Eq.(1.22)),
we get the zero Nijenhuis tensor condition of
$$
N(X,Y)=0\eqno(2.16)
$$
for any
$X,Y\in W$
and vice versa.
Moreover,
$J$ satisfies also the
analogue of the almost complex
structure condition 
of
$$
J^{2}=
-
\left(
\begin{array}{cc}
1&0\\
0&1
\end{array}
\right).
\eqno(2.17)
$$
\par
We remark that these facts
are already noted in
Proposition 5.3
of
[K-O]
for the special case of 
$\varepsilon=\delta.$
Further for
$\varepsilon=\delta$,
$J$ is a derivation
(as well as an automorphism)
of the Lie algebra or superalgebra
$\hat{L}$,
and we may
replace
$J$ by
$$
J\rightarrow \tilde{J}=UJU^{-1}\eqno(2.18)
$$
for any
$2\times 2$ matrix
$U$
satisfying 
${\rm Det}\ U=1$
by the following reason.
First,
we see that
Eqs(2.13)
with 
(2.16) is invariant under
$SL(2)$
transformation of
$$
J\rightarrow \tilde{J}=
UJU^{-1},\ 
X\rightarrow
\tilde{X}=UX,\ 
Y\rightarrow
\tilde{Y}=UY.$$
\par
Moreover,
since
$X$ and
$Y$ are arbitrary,
we may replace 
$X$ and $Y$ by
$U^{-1}X$
and $U^{-1}Y$,
respectively,
to obtain the desired result,
i.e..
$$
\tilde{N}(X,Y)=
[\tilde{J}X,\tilde{J}Y]-
\tilde{J}[\tilde{J}X,Y]-
\tilde{J}
[X,\tilde{J}Y]+
\tilde{J}^{2}
[X,Y]=0,\eqno(2.19)
$$
with
$$
\tilde{J}^{2}=-
\left(
\begin{array}{cc}
1&0\\
0&1
\end{array}
\right).
$$
\par
A simple example of
a special $(\varepsilon,\delta )$ FKTS is
given as follows.
Let $<\ \cdot | \ \cdot>$  be a bilinear form
in a vector space $V$, satisfying
$<x|y>=-\varepsilon <y|x>$,
and define a tri-linear product $xyz$ in $V$ by
$$
xyz=<y|z>x.
$$
Then $(V,\ xyz)$ is
a special $(\varepsilon,\delta)$FKTS.
The fact that it gives a $(\varepsilon,\delta)$ FKTS
has been already noted in
([K-O Proposition 2.8 (ii)]). 
In order to show it
to be special,
we calculate
$$
K(x,y)z=xzy-\delta yzx=<z|y>x-\delta <z|x>y
$$
$$
=-\varepsilon <y|z>x +\varepsilon \delta <x|z>y=-\varepsilon xyz + \varepsilon \delta yxz
$$

$$=
-\varepsilon L(x,y)z+ \varepsilon \delta L(y,x)z
$$
and hence $K(x,y)=-\varepsilon L(x,y) +\varepsilon \delta L(y,x)$.
\par
{\bf Remark 2.6}
\par
Let $\hat{ L}=W\oplus L(W,W)$ be the 
Lie algebra derived from a Lie triple system $[W,W,W]$,
and introduce an analogue
of covariant derivative $\nabla : \hat{ L}\rightarrow End \hat{ L}$
by
$$
\nabla _{X}Y =[X,Y],\ \ \ \
\nabla _{X}[Y,Z]=[Y,Z,X],
$$
$$
\nabla _{[X,Y]}Z=-[X,Y,Z],\ \ \ \
\nabla_{[X,Y]}[V,Z]=-[[X,Y],[V,Z]].
$$
Then
the Riemann curvature tensor defined by (see [K-N]) 
$$
R(X,Y)=\nabla _{X}\nabla _{Y}-\nabla _{Y}\nabla _{X}-\nabla _{[X,Y]}
$$
is identically zero,
i.e.,
$R(X,Y)=0$
in $\hat{ L}$,
as we demonstrate below.
First we calculate
$$
R(X,Y)Z=(\nabla _{X}\nabla _{Y}-\nabla _{Y}\nabla _{X})Z-\nabla _{[X,Y]}Z
$$
$$
=\nabla _{X}[Y,Z]-\nabla _{Y}[X,Z]+[X,Y,Z]=[Y,Z,X]-[X,Z,Y]+[X,Y,Z]
$$
$$
=[Y,Z,X]+[Z,X,Y]+[X,Y,Z]=0.
$$
Second,
$$
R(X,Y)[V,Z]=(\nabla _{X}\nabla _{Y}-\nabla _{Y}\nabla _{X})[V,Z]
-\nabla _{[X,Y]}[V,Z]
$$
$$
= [X,[V,Z,Y]]-[Y,[V,Z,X]]+[[X,Y],[V,Z]]
$$
$$
=[X,L(V,Z)Y]-[Y,L(V,Z)X]-L(V,Z)[X,Y]=0
$$
where
$L(W,W) $
is defined as
before by
$$
L(X,Y)Z=[X,Y,Z],
$$
and
we note that
it is
a derivation of the Lie triple system.
\par
However the torsion tensor $T(X,Y)$
defined by 
$$
T(X,Y)= \nabla _{X}Y-\nabla _{Y}X -[X,Y]
$$
is not zero,
since it gives
$$
T(X,Y)=[X,Y]-[Y,X]-[X,Y]=[X,Y].
$$
In conclusion, 
we see that the Lie
triple system
associated with the $(\varepsilon, \delta)$ FKTS 
contains many interesting structures in it.
\par
\vskip 3mm
{\bf 3.\ Structurable algebras and $S_{4}$-
symmetry}
\par
\vskip 3mm
Let $A=(V,xy)$
be a structurable algebra with the
unit element
$e$ and with involution map
$x \rightarrow \bar{x}$.
Let
$l(x)$ and $r(x)$ be the left and
right multiplication operators defind by
$$
l(x)y=xy,\quad r(x)y=yx,\eqno(3.1)
$$
we introduce then
$d_{j}:A\otimes A\rightarrow
{\rm End}A$
for
$j=1,2,3$
by
$$
\begin{array}{ll}
d_{1}(x,y)
&=
l({\bar y})l(x)-
l({\bar x})l(y)
\hskip 95mm
(3.2a)\\
d_{2}(x,y)
&=
r({\bar y})r(x)-
r({\bar x})
r(y)
\hskip 93mm
(3.2b)\\
d_{3}(x,y)
&=
r({\bar x}y-{\bar y}x)+
l(y)l({\bar x})-
l(x)l({\bar y})\hskip 70mm
(3.2c)\\
&
=l(y{\bar x}-x{\bar y})+
r(y)
r({\bar x})-
r(x)
r({\bar y}),
\end{array}
$$
following [A-F].
Note that they satisfy
$$
d_{j}(x,y)=-d_{j}(y,x),\ 
j=1,2,3.\eqno(3.3)
$$
\par
It is known ([A-F],[O])
then that they satisfy first the
triality
relation:
$$
\overline{d_{j}(u,v)}(xy)=
(d_{j+1}(u,v)x)y+
x(d_{j+2}(u,v)y)\eqno(3.4)
$$
for any $u,v,x,y\in A$
and
$j=1,2,3$.
Here,
$d_{j}(u,v)$
is defined modulo
$3$ 
for the
index
$j$,
i.e..
$$
d_{j\pm 3}(u,v)=
d_{j}(u,v)$$
and
$\bar{Q}\in {\rm End}\ A$
for any
$Q\in {\rm End}\ A$ is defined as 
usual by
$$
\overline{Qx}=
\bar{Q}\bar{x}.
$$
\par
Moreover,
they satisfy
$$
\begin{array}{ll}
(i)&
d_{3}(x,y)z+
d_{3}(y,z)x+
d_{3}(z,x)y=0
\hskip 65.5mm
(3.5a)\\
(ii)&
d_{1}({\bar x},yz)+d_{2}(\bar{y},zx)+
d_{3}(\bar{z},xy)=0
\hskip 65.5mm
(3.5b)\\
(iii)&
\overline{d_{j}}(x,y)=d_{3-j}(\bar{x},\bar{y})
\hskip 90mm
(3.5c)\\
(iv)&
[d_{j}(u,v),d_{k}(x,y)]=
d_{k}(d_{j-k}(u,v)x,y)+
d_{k}(x,d_{j-k}(u,v)y).\hskip 26.5mm
(3.5d)
\end{array}
$$
Conversely,
the validity of Eqs.(3.4) and
(3.5) imply that the algebra is structurable,
if it is unital and involutive.
\par
Althogh the structurable algebra is intimately
related to
$(-1,1)$FKTS,
we can construct another type of  Lie algebra
out of it,
independently of the standard construction
given in section 1 as follows:
Let $\rho_{j}(A)$ for
$j=1,2,3$
be 3 copies of $A$,
and we introduce symbols
$T_{j}(A,A)$ satisfying
$$
\begin{array}{ll}
(i)&
T_{j}(x,y)=
-
T_{j}(y,x)=
T_{j\pm 3}(x,y)
\hskip 78.5mm
(3.6a)\\
(ii)&
T_{1}(\bar{x},yz)+
T_{2}({\bar y},zx)+
T_{3}({\bar z},xy)=0.
\hskip 72mm
(3.6b)
\end{array}
$$
Let $T(A,A)$ be a linear span of all
$T_{j}(x,y)$
for $x,y\in A$
and consider
$$
L=
\rho_{1}(A)\oplus
\rho_{2}(A)\oplus
\rho_{3}(A)\oplus
T(A,A).\eqno(3.7)
$$
Then,
$L$ is a Lie algebra 
(see
[A-F],
and [O])
with Lie brackets of
$$
\begin{array}{ll}
(i)&
[\rho_{i}(x),\rho_{i}(y)]=
\gamma_{j}\gamma_{k}^{-1}T_{3-i}(x,y)
\hskip 82mm
(3.7a)\\
(ii)&
[\rho_{i}(x),\rho_{j}(y)]=
-[\rho_{j}(y),\rho_{i}(x)]=
-\gamma_{j}\gamma_{i}^{-1}\rho_{k}
(\overline{xy})
\hskip 54mm
(3.7b)\\
(iii)&
[T_{l}(u,v),\rho_{j}(x)]=
-[\rho_{j}(x),T_{l}(u,v)]=
\rho_{j}(d_{j+l}(u,v)x)
\hskip 45.5mm
(3.7c)\\
(iv)&
[T_{l}(u,v),T_{m}(x,y)]=\\
&
-[T_{m}(x,y),T_{l}(u,v)]=
T_{m}(x,d_{l-m}(u,v)y)+
T_{m}(d_{l-m}(u,v)x,y).
\hskip 26.5mm
(3.7d)
\end{array}
$$
Here
$(i,j,k)$
is any cyclic permutation of indices
$(1,2,3)$
with
$\gamma_{j}'s$ being any non-zero
 constants,
 while indicies $l$ and $m$ are arbitrary integers
 and we assumed
 $\rho_{j}(x)$ to be $F$-linear in $x$.
 A economical choice for
 $T_{l}(x,y)$
 is to assume it to be a triple
 $$
 T_{l}(x,y)=
 T(d_{l}(x,y),
 d_{l+1}(x,y),
 d_{l+2}(x,y))$$
 as in [A-F] and [E],
 since the Eqs.(3.6)
 and (3.7d) are automatically
 satisfied by
 Eqs.(3.5b) and (3.5d).
 However,
 this choice
 is not suitable in what follows.
 \par
 A special choice of
 $\gamma_{1}=\gamma_{2}=\gamma_{3}=1$
 is of particular interest,
 since the Lie algebra
 $L$ is then invariant under $S_{4}$-symmetry
 as follows:
First,
$L$ is invariant under the cyclic permutation group $Z_{3}$
generated by the permutation
$(1,2,3)$,
i.e.,
$1\rightarrow 2\rightarrow 3\rightarrow 1$ by
$$
\rho_{j}(x)\rightarrow \rho_{j+1}(x),\quad
T_{j}(x,y)\rightarrow T_{j-1}(x,y).\eqno(3.8)$$
\par
The action of
$\tau=(1,2)$ of the
$S_{3}$-group is given by
$$
\rho_{1}(x)
\leftrightarrow -\rho_{2}(\bar{x}),\quad
\rho_{3}(x)\rightarrow
-\rho_{3}(\bar{x})\eqno(3.9a)$$
$$
T_{1}(x,y)\leftrightarrow T_{2}(\bar{x},\bar{y}),\quad
T_{3}(x,y)\rightarrow
T_{3}(\bar{x},\bar{y}).\eqno(3.9b)
$$
Then,
$S_{3}$-group generated by $(1,2,3)$
and
$\tau =(1,2)$
can be shown to be
automorphism of the Lie algebra $L$.
Next,
we consider the Klein's
$4$-group 
$K_{4}$
conrresponding to permutations
$$
\tau_{1}=
(2,3)(1,4),
\quad
\tau_{2}=(1,3)(2,4),\quad
\tau_{3}=
(1,2)(3,4)\eqno(3.10)$$
which satisfy
$$
\tau_{i}\tau_{j}=
\tau_{j}\tau_{i},\quad
\tau_{i}\tau_{i}=1,\quad
\tau_{1}\tau_{2}\tau_{3}=1\eqno(3.11)
$$
for
$i,j=1,2,3.$
The actions of $\tau_{1}$
for example for
 $L$ is then realized by
$$
\tau_{1}:
\rho_{1}(x)\rightarrow \rho_{1}(x),\ 
\rho_{2}(x)\rightarrow -\rho_{2}(x),\ 
\rho_{3}(x)\rightarrow -\rho_{3}(x),\eqno(3.12a)$$
$$
T_{i}(x,y)\rightarrow T_{i}(x,y)$$
and similarly for
$\tau_{2}$ and
$\tau_{3}$.
Then,
again we see that
$L$ is invariant under this
Klein's $4$-group.
Since $S_{4}$ can be generated by
$S_{3}$ and
$K_{4}$,
this shows that $L$ is invariant under
$S_{4}$ as has been noted in
[E-O,1].
\par
Now, a question arises about relations between this construction
 of Lie
algebra and that based upon the standard construction 
from
$(-1,1)$FKTS as in section 1.
Note that the symmetry group $D(-1,1)$
admites $S_{3}$-symmetry but not $S_{4}$.
The purpose of this section is to show that the Lie algebra
$L$ given in this section can be obtained from that of
$\hat{L}$ constructed in section 1 by prolonging
$\hat{L}$ when we 
take note of $A$ to be structurable,
provided that the underlying
field $F$ contains the square root 
$\sqrt{-1}$
of $-1$.
More precisely,
we will prove the following theorem.
\par
\vskip 3mm
{\bf Theorem 3.1}
\par
\vskip 3mm
\it
Let $A=(V,xy)$ be a structurable algebra.
Let
$\hat{L}=L(W,W)\oplus W$
be the Lie algebra constructed as in section 1
from the associated
$(-1,1)$FKTS.
First,
for any non-zero constants
$\alpha,\beta,k\in F$,
we introduce
the ratio
$\frac{\gamma_{1}}{\gamma_{3}}$ 
and $\frac{\gamma_{2}}{\gamma_{3}}$
for some
$\gamma_{j}\in F$
by
$$
\begin{array}{ll}
(i)&
\frac{\gamma_{2}}{\gamma_{3}}=-2\alpha\beta
\hskip 106.5mm
(3.13a)\\
(ii)&
\frac{
(\gamma_{3})^{2}}{\gamma_{1}\gamma_{2}}=
-k^{2}
\hskip 106mm
(3.13b)
\end{array}
$$
and second
define
$\rho_{j}(x)$
and
$T_{j}(x,y)$
in $\hat{L}=L(W,W)\oplus W$
by
$$
\begin{array}{ll}
(1)&
\rho_{1}(x)=
\left(
\begin{array}{c}
\alpha x\\
\beta x
\end{array}
\right)\hskip 105mm
(3.14a)\\
(2)&
\rho_{2}(x)=
\left(
\begin{array}{c}
k\alpha{\bar x}\\
-k\beta{\bar x}
\end{array}
\right)
\hskip 100.5mm
(3.14b)\\
(3)&
\rho_{3}(x)=
k
(\frac{\gamma_{1}}{\gamma_{2}})
\left(
\begin{array}{cc}
\alpha\beta l(x+{\bar x}),&
\alpha^{2}l(x-{\bar x})\\
-\beta^{2}l(x-{\bar x}),&
-\alpha\beta l(x+{\bar x})
\end{array}
\right)
\hskip 52mm
(3.15c)\\
(4)&
T_{1}(x,y)=\frac{\gamma_{3}}{\gamma_{2}}
\left(
\begin{array}{cc}
\alpha\beta(L({\bar x},{\bar y})-L({\bar y},{\bar x})),&
+\alpha^{2}K({\bar x},{\bar y})\\
-\beta^{2}K({\bar x},{\bar y}),&
\alpha\beta(L({\bar x},{\bar y})-
L({\bar y},{\bar x}))
\end{array}
\right)
\hskip 25mm(3.15d)\\
(5)&
T_{2}(x,y)=\frac{\gamma_{3}}{\gamma_{2}}
\left(
\begin{array}{cc}
\alpha \beta(L(x,y)-L(y,x)),&
-\alpha^{2}K(x,y)\\
\beta^{2}K(x,y),&
\alpha\beta(L(x,y)-
L(y,x))
\end{array}
\right)
\hskip 27mm
(3.15e)\\
(6)&
T_{3}(x,y)=\frac{\gamma_{2}}{\gamma_{1}}
[\rho_{3}(x),\rho_{3}(y)].
\hskip 92.5mm
(3.15f)
\end{array}
$$
Then,
$\rho_{j}(x)$
and $T_{j}(x,y)$
are elements of
$L(W,W)\oplus W$
and satisfy the Lie algebra
relation of Eqs.(3.7) and (3.6).
\par
\vskip 3mm
{\bf Remark 3.2}\par
\vskip 3mm
\rm
It is not self-evident that
$\rho_{3}(x)$
and $T_{3}(x,y)$
are elements of
$L(W,W)\oplus W.$
First,
as we will show soon,
we have
$$
l(x+{\bar x})=
L(e,x)+L(x,e)\eqno(3.16a)$$
$$
l(x-{\bar x})=
K(x,e)=-K(e,x)\eqno(3.16b)
$$
which prove
$\rho_{3}(x)\in L(W,W),$
where $e$ is the unit element of 
$A$.
Then this also implies
$T_{3}(x,y)$
to be an element of
$L(W,W)$
since
$\hat{L}=L(W,W)\oplus W$
is a
$Z_{2}$-graded Lie algebra as we noted in section 1.
\par
Before going into a proof of Theorem 3.1,
we make
a comment on the
$S_{4}$-symmetry of the Lie algebra constructed here.
The special choice of
$\gamma_{1}=\gamma_{2}=\gamma_{3}=1$
requires
$k^{2}=-1$
and
$2\alpha\beta=-1$
by Eqs.(3.13)
so that we must assume $F$
to contain the
element
$\sqrt{-1}.$
\par
We now proceed for a proof
of Theorem 3.1.
First,
we show 
\par
\vskip 3mm
{\bf Lemma 3.3}
\par
\vskip 3mm
\it
Under the assumption as in above,
we have
$$
\begin{array}{ll}
(1)&
L(x,y)+L(y,x)=
l(x{\bar y}+
y{\bar x})\hskip 85mm
(3.17a)\\
(2)&
L(x,y)-L(y,x)=
-d_{2}({\bar x},{\bar y})-
d_{0}(x,y)
\hskip 70mm
(3.17b)\\
(3)&
K(x,y)=
d_{2}({\bar x},{\bar y})-
d_{0}(x,y)=
l(x{\bar y}-y{\bar x}).\hskip 62.5mm
(3.17c)
\end{array}
$$ 
\par
\vskip 3mm
{\bf Proof}
\par
\vskip 3mm
\rm
Since
$xyz=(x{\bar y})z-
(z{\bar x})y+
(z{\bar y})x,$
this is rewritten as
$$
L(x,y)=
r(x)r({\bar y})-
r(y)r({\bar x})+
l(x{\bar y}).$$
\par
On the other side,
we note
$$
\begin{array}{ll}
K(x,y)z
&
=
xzy-yzx=
(y{\bar z})x-
(y{\bar x})z+
(x{\bar z})y-
(x{\bar z})y+
(x{\bar y})z-
(y{\bar z})x\\
&=
(x{\bar y}-y{\bar x})z=
l(x{\bar y}-y{\bar x})z=
\{d_{2}({\bar x},{\bar y})
-d_{0}(x,y)\}z.
\end{array}
$$
Then,
Eqs.(3.17)
follow readily from these relations.
Setting
$y=e$ in Eq.(3.17a)
gives Eq.(3.16a),
while Eq.(3.16b)
is a simple consequence of
Eq.(3.17c) for
$y=e.$
This completes the proof.//
\par
 Following ([KMO]),
 we note that
\par
$
D_{x,y}:=
L(x,y)-L(y,x)$ is a derivation of $A$
with respect to (abbreviation hereafter as to w.r.t)
the triple product
$(xyz)$,
\par
$A_{x,y}:=
L(x,y)+L(y,x)$
is an anti- derivation
of
$A$ w.r.t
the triple product
$(xyz)$
i.e.,
$$
[D_{x,y},L(a,b)]=
L(D_{x,y}a,b)+
L(a,D_{x,y}b)
$$
$$
[A_{x,y},L(a,b)]=
L(A_{x,y}a,b)-
L(a,A_{x,y}b),
$$
and furthermore,
note
\par
$[K(x,y),K(a,b)]$
is a derivation of
$A$ w.r.t.
the triple product, 
\par
$[A_{x,y},A_{a,b}]$
is a derivation of
$A$ w.r.t.
the triple product,
and
\par
$[A_{x,y}, D_{a,b}]$
is an anti-derivation of $A$
w.r.t.
the triple system.
\par
\vskip 3mm
{\bf Lemma 3.4}
\par
\vskip 3mm
\it
Under the assumption as in above,
we write
$$
[X,Y]=
XY-YX,\ {\rm and}\ 
\{X,Y\}_{+}=
XY+YX\ {\rm for}\ 
X,Y\in 
\ {\rm End}\ V.$$
\par
We then have
$$
\begin{array}{ll}
(1)&
[d_{0}({\bar x},{\bar y})+
d_{2}(x,y),
l(z)]=
l((d_{0}(x,y)+
d_{2}({\bar x},{\bar y}))z)\hskip 50mm
(3.18a)\\
(2)&
\{d_{0}({\bar x},{\bar y})-
d_{2}(x,y),
l(z)\}_{+}
=
l((d_{0}(x,y)-
d_{2}({\bar x},{\bar y}))z)\hskip 45.5mm
(3.18b)\\
(3)&
[d_{1}(x,y)+
d_{1}({\bar x},{\bar y}),
l(z)]=
l((d_{0}
({\bar x,}{\bar y})+
d_{0}(x,y)z)\hskip 52.5mm
(3.18c)\\
(4)&
\{d_{1}(x,y)-
d_{1}({\bar x},{\bar y}),
l(z)\}_{+}=
l((d_{0}({\bar x},{\bar y})-
d_{0}(x,y)z).\hskip 44.5mm
(3.18d)
\end{array}
$$
\par
\vskip 3mm
{\bf Proof}
\par
\vskip 3mm
\rm
We rewrite the triality relation
Eq.(3.4) as
$$
d_{3-j}({\bar x},{\bar y})(zw)=
(d_{j+1}(x,y)z)w+
z(d_{j+2}(x,y)z)$$
by Eq.(3.5c),
which gives
$$
d_{3-j}({\bar x},{\bar y})l(z)=
l(d_{j+1}(x,y)z)+
l(z)d_{j+2}( x, y).
$$
Letting $x\rightarrow {\bar x}$
and
$y\rightarrow {\bar y}$
with
$j\rightarrow 1-j,$
we have also
$$
d_{j+2}(x,y)l(z)=
l(d_{2-j}({\bar x},{\bar y})z)+
l(z)
d_{3-j}({\bar x},{\bar y}).
$$
Adding and for subtracting both relatians,
we obtain
$$
\begin{array}{l}
[d_{3-j}({\bar x},{\bar y})+
d_{j+2}(x,y),
l(z)]=
l((d_{j+1}(x,y)+
d_{2-j}({\bar x},{\bar y}))z)\\
\{d_{3-j}({\bar x},{\bar y})-d_{j+2}(x,y),
l(z)\}_{+}=
l((d_{j+1}(x,y)-
d_{2-j}({\bar x},{\bar y}))z).
\end{array}
$$
Setting
$j=2\ {\rm and}\ 3$,
these give Eqs.(3.18).
This completes the proof.//
\par
\vskip 3mm
{\bf Lemma 3.5}
\par
\vskip 3mm
\it
Under the assumptions as in above,
we have
$$
T_{j}(x,y)=
-\frac{\gamma_{3}}{\gamma_{2}}
\left(\begin{array}{cc}
\alpha\beta
(d_{j+1}(x,y)+
d_{1-j}({\bar x},{\bar y})),&
\alpha^{2}(d_{j+1}(x,y)-
d_{1-j}({\bar x},{\bar y}))\\
\beta^{2}
(d_{j+1}(x,y)-
d_{1-j}({\bar x},{\bar y})),&
\alpha\beta
(d_{j+1}(x,y)+
d_{1-j}({\bar x},{\bar y}))
\end{array}
\right)
\eqno(3.19)
$$
for
$j=1,2,3.$
\par
\vskip 3mm
{\bf Proof}
\par
\vskip 3mm
\rm
The case of
$j=1$ and
$2$ are simple rewriting of
Eqs.(3.15d)
and (3.15e)
together with Eq.(3.17).
In order to prove it for
$j=3$,
we calculate
$$
\begin{array}{ll}
T_{3}(x,y)&
=
\frac{\gamma_{2}}{\gamma_{1}}
\{\rho_{3}(x)
\rho_{3}(y)-
\rho_{3}(y)
\rho_{3}(x)\}\\
&=
(\frac{\gamma_{2}}{\gamma_{1}})
(\frac{\gamma_{1}}{\gamma_{2}})^{2}
k^{2}
[
\left(
\begin{array}{cc}
\alpha\beta l(x+{\bar x}),&
\alpha^{2}l(x-{\bar x})\\
-\beta^{2}l(x-{\bar x}),&
-\alpha\beta l(x+{\bar x})
\end{array}
\right),
\left(
\begin{array}{cc}
\alpha\beta l(y+{\bar y}),&
\alpha^{2} l(y-{\bar y})\\
-\beta^{2}l(y-{\bar y}),&
\alpha\beta l({\bar y}+{\bar y})
\end{array}
\right)]\\
&=
k^{2}
\frac{\gamma_{1}}{\gamma_{2}}
\alpha\beta
\left(
\begin{array}{cc}
\alpha\beta A(x,y)&
\alpha^{2}B(x,y)\\
\beta^{2} B(x,y)&
\alpha\beta A(x,y)
\end{array}
\right)
\end{array}
$$
where we have set
$$
\begin{array}{ll}
A(x,y)&
=
[l(x+{\bar x}),
l(y+{\bar y})]-
[l(x-{\bar x}),l(y-{\bar y})]\\
&=
2\{
[l(x),l({\bar y})]+
[l({\bar x}),l(y)]\}=
-2\{d_{1}(x,y)+
d_{1}({\bar x},{\bar y})\}
\end{array}
$$
and
$$
\begin{array}{ll}
B(x,y)&=
\{l(x+{\bar x}), l(y,{\bar y})\}_{+}-
\{l(x-{\bar x}),
l(y+{\bar y})\}_{+}\\
&=
2(\{l({\bar x}),
l(y)\}_{+}-
\{l(x),
l({\bar y})\}_{+})=
2(d_{1}({\bar x},{\bar y})-
d_{1}(x,y))
\end{array}$$
from  Eqs.(3.2).
Also,
we note
$$
k^{2}
\frac{\gamma_{1}}{\gamma_{2}}\alpha\beta=
\frac{(\gamma_{3})^{2}}{\gamma_{1}\gamma_{2}}
(\frac{\gamma_{1}}{\gamma_{2}})
\frac{\gamma_{2}}{2\gamma_{3}}=
\frac{\gamma_{3}}{2\gamma_{2}}
$$
so that these yield
$$
T_{3}(x,y)=
-\frac{\gamma_{3}}{\gamma_{2}}
\left(
\begin{array}{cc}
\alpha\beta(d_{1}(x,y)+
d_{1}({\bar x},{\bar y})),&
\alpha^{2}
(d_{1}(x,y)-
d_{1}({\bar x},{\bar y}))\\
\beta^{2}(d_{1}(x,y)-
d_{1}({\bar x},{\bar y})),&
\alpha\beta(d_{1}(x,y)+
d_{1}({\bar x},{\bar y}))
\end{array}
\right).
$$
This completes the proof of Lemma 3.5.//
\par
\vskip 3mm
{\bf Remark 3.6}
\par
\vskip 3mm
Eq.(3.19) together with
Eqs.(3.5) prove the validity of
Eq.(3.6b),i.e.,
$$
T_{1}({\bar x},yz)+
T_{2}({\bar y},zx)+
T_{3}({\bar z},xy)=0.$$
Also,
it is still not a trivial matter to see that
$T_{3}(x,y)\in L(W,W).$
To show it,
we note
$$
\begin{array}{l}
d_{1}(x,y)-d_{1}({\bar x},{\bar y})=
-K({\bar x},{\bar y})+K(x,y)\hskip 75mm
(3.20a)\\
d_{1}(x,y)+
d_{1}({\bar x},{\bar y})=
L(y,x)-L(x,y)+
L(e,{\bar x}y)-
L({\bar x}y,e).\hskip 40mm
(3.20b)
\end{array}
$$
Since by putting
$z=e$
 and
$y=e$
in Eq.(3.5b) and 
changing
notations suitably,
we obtain
$$
d_{1}(x,y)=
-d_{3}(e,{\bar x}y)+
d_{2}({\bar x},{\bar y})=
d_{3}({\bar x},{\bar y})-
d_{2}(e,y{\bar x})$$
and hence
$$
d_{1}(x,y)-
d_{1}({\bar x},{\bar y})=
d_{2}({\bar x},{\bar y})-
d_{3}(x,y)-
d_{3}(e,{\bar x}y)+
d_{2}(e,{\bar y}x)=
K(x,y)+
K(e,{\bar x}y).$$
However,
$$
K(e,{\bar x}y)=
l(e(\overline{{\bar x}y})-
({\bar x}y)e)=
l({\bar y}x-{\bar x}y)=
-K({\bar x},{\bar y}),
$$
and also
$$
\begin{array}{ll}
d_{1}(x,y)+
d_{1}({\bar x},{\bar y})&
=d_{2}({\bar x},{\bar y})+
d_{3}(x,y)-
d_{3}(e,{\bar x}y)-
d_{2}(e,{\bar y}x)\\
&
=-\{L(x,y)-L(y,x)-
L(e,{\bar x}y)+
L({\bar x}y,e)\}.
\end{array}
$$
\par
We next show
\par
\vskip 3mm
{\bf Proof of Eq.(3.7a)}
\par
\vskip 3mm
The case of 
$j=3$ is trivial in view of 
Eq.(3.15f)
\par
(i)\ For $j=1,$
we calcutate
$$
\begin{array}{ll}
[\rho_{1}(x),\rho_{1}(y)]
&
=[\left(\begin{array}{c}
\alpha x\\
\beta x
\end{array}
\right),
\left(
\begin{array}{c}
\alpha y\\
\beta y
\end{array}
\right)]\\
&=
\left(
\begin{array}{cc}
L(\alpha x,\beta y)-L(\alpha y,
\beta x) ,&
K(\alpha x,\alpha y)\\
K(\beta x,\beta y),&
L(\beta x,\alpha y)-L(\beta y,\alpha x)
\end{array}
\right)\\
&=
\left(\begin{array}{cc}
\alpha\beta(L(x,y)-L(y,x)),&
\alpha^{2} K(x,y)\\
\beta^{2} K(x,y),&
\alpha\beta(L(x,y)-
L(y,x))
\end{array}
\right)=
\frac{\gamma_{2}}{\gamma_{3}}
T_{2}
(x,y).
\end{array}
$$
\par
(ii)\ Similarly,
we compute for
$j=2$
$$
\begin{array}{ll}
[\rho_{2}(x),
\rho_{2}(y)]&
=k^{2}[
\left(\begin{array}{c}
\alpha{\bar x}\\
-\beta{\bar x}
\end{array}
\right),
\left(
\begin{array}{c}
\alpha{\bar y}\\
-\beta{\bar y}
\end{array}
\right)]\\
&=
k^{2}
\left(\begin{array}{cc}
L(\alpha{\bar x},-\beta{\bar y})-L(\alpha {\bar y},-\beta{\bar x}),&
K(\alpha{\bar x},\alpha{\bar y})\\
K(-\beta{\bar x},-\beta{\bar y}),&
L(-\beta{\bar x},\alpha{\bar y})-
L(-\beta{\bar y},\alpha{\bar x})
\end{array}
\right)\\
&=
-k^{2}
(\frac{\gamma_{2}}{\gamma_{3}})
T_{1}(x,y)=
\frac{\gamma_{3}}{\gamma_{1}}
T_{1}(x,y).
\end{array}
$$
\par
This completes the proof of (3.7a).
\vskip 3mm
{\bf Proof of Eq.(3.7b)}
\par
\vskip 3mm
(i)
We calculate
$$
\begin{array}{ll}
[\rho_{1}(x),\rho_{2}(y)]&
=[\left(
\begin{array}{c}
\alpha x\\
\beta x
\end{array}
\right),
\left(
\begin{array}{c}
k\alpha{\bar x}\\
-k\beta{\bar y}
\end{array}
\right)]\\
&=
k
\left(\begin{array}{cc}
L(\alpha x,-\beta{\bar y})
-L(\alpha{\bar y},\beta x),&
K(\alpha x,\alpha{\bar x})\\
K(\beta x,-\beta{\bar y}),&
L(\beta x,\alpha{\bar y})
-L(-\beta{\bar y},
\alpha x)
\end{array}
\right)\\
&=
-k
\left(\begin{array}{cc}
\alpha\beta(L(x,{\bar y})+
L({\bar y},x)),&
-\alpha^{2}K(x,{\bar y})\\
\beta^{2}K(x,{\bar y}),&
-\alpha\beta(L(x,{\bar y})+
L({\bar y},x))
\end{array}
\right)\\
&=
-k
\left(
\begin{array}{cc}
\alpha\beta l(xy+{\bar y}{\bar x}),&
-\alpha^{2}l
(xy-{\bar y}{\bar x})\\
\beta^{2}l(xy-{\bar y}{\bar x}),&
-\alpha\beta l(xy+
{\bar y}{\bar x})\end{array}
\right)\\
&=
-\frac{\gamma_{2}}{\gamma_{1}}
\rho_{3}
({\bar y}{\bar x})=
-\frac{\gamma_{2}}{\gamma_{1}}
\rho_{3}
(\overline{xy}),
\end{array}$$
\par
(ii)
Similerly we have
$$
\begin{array}{ll}
[\rho_{2}(x),\rho_{3}(y)]&
=
-[\rho_{3}(y),\rho_{2}(x)]\\
&
=-k^{2}
(\frac{\gamma_{1}}{\gamma_{2}})
\left(
\begin{array}{cc}
\alpha\beta l(y+{\bar y}),&
\alpha^{2}l(y-{\bar y})\\
-\beta^{2}l(y-{\bar y}),&
-\alpha\beta l(y+{\bar y})
\end{array}\right)
\left(
\begin{array}{c}
k\alpha{\bar x}\\
-k\beta{\bar x}
\end{array}
\right)\\
&
=
-k^{2}
(\frac{\gamma_{1}}{\gamma_{2}})
\left(
\begin{array}{c}
(\alpha^{2}\beta(y+{\bar y})
-\alpha^{2}\beta(y-{\bar y})){\bar x}\\
(-\beta^{2}\alpha(y-{\bar y})+
\beta^{2}\alpha(y+{\bar y}))
{\bar x}
\end{array}
\right)\\
&
=
-k^{2}
(\frac{\gamma_{1}}{\gamma_{2}})
\left(
\begin{array}{c}
2\alpha^{2}\beta{\bar y}{\bar x}\\
2\beta^{2}\alpha{\bar y}{\bar x}
\end{array}
\right)\\
&=
-2k^{2}
(\frac{\gamma_{1}}{\gamma_{2}})
\alpha\beta
\left(\begin{array}{c}
\alpha{\bar y}{\bar x}\\
\beta
{\bar y}{\bar x}
\end{array}
\right)=
-
\frac{\gamma_{3}}{\gamma_{2}}
\rho_{1}
(\overline{xy})
\end{array}
$$
\par
(iii)
Similarly we have
$$
\begin{array}{ll}
[\rho_{3}(x),\rho_{1}(y)]&
=
k(
\frac{\gamma_{1}}{\gamma_{2}})
\left(
\begin{array}{cc}
\alpha\beta l(x+{\bar x}),&
\alpha^{2}l(x-{\bar x})\\
-\beta^{2}l(x-{\bar x}),&
-\alpha\beta l(x+{\bar x})
\end{array}
\right)
\left(
\begin{array}{c}
\alpha y\\
\beta y
\end{array}
\right)\\
&=
k(\frac{\gamma_{1}}{\gamma_{2}})
\left(
\begin{array}{c}
\alpha^{2}\beta(x+{\bar x})y+
\alpha^{2}\beta(x-{\bar x})y\\
-\beta^{2}\alpha(x-{\bar x})y-
\alpha\beta^{2}(x+{\bar x})y
\end{array}
\right)\\
&=
k(\frac{\gamma_{1}}{\gamma_{2}})
\left(
\begin{array}{c}
2\alpha^{2}\beta xy\\
-2\beta^{2}\alpha xy
\end{array}
\right)\\
&=
2(\frac{\gamma_{1}}{\gamma_{2}})
\alpha\beta\rho_{2}
(\overline{xy})=
-\frac{\gamma_{1}}{\gamma_{2}}
\rho_{2}(\overline{xy}).
\end{array}
$$
This complete the proof of Eq.(3.7b).
\par
\vskip 3mm
{\bf Proof of Eq.(3.7c)}
\par
\vskip 3mm
(1)\ The case of $j=1,$
we have
\par
$$
\begin{array}{l}
[T_{l}(x,y),\rho_{1}(z)]\\
=
-\frac{\gamma_{3}}{\gamma_{2}}
\left(
\begin{array}{cc}
\alpha\beta(d_{l+1}(x,y)+
d_{1-l}({\bar x},{\bar y})),&
\alpha^{2}(d_{l+1}(x,y)-
d_{1-l}({\bar x},{\bar y}))\\
\beta^{2}(d_{l+1}(x,y)-
d_{1-l}(
{\bar x},{\bar y})),&
\alpha\beta(d_{l+1}(x,y)+
d_{1-l}({\bar x},{\bar y}))
\end{array}
\right)
\left(
\begin{array}{c}
\alpha z\\
\beta z
\end{array}
\right)\\
=
-\frac{\gamma_{3}}{\gamma_{2}}
\left(
\begin{array}{l}
\alpha^{2}\beta(d_{l+1}(x,y)+
d_{1-l}({\bar x},{\bar y}))z+
\alpha^{2}\beta
(d_{l+1}(x,y)-
d_{1-l}
({\bar x},{\bar y}))z\\
\beta^{2}\alpha
(d_{l+1}(x,y)-
d_{1-l}({\bar x},{\bar y}))z+
\beta^{2}\alpha
(d_{l+1}
(x,y)+
d_{1-l}
({\bar x},{\bar y}))z
\end{array}
\right)\\
=
-2\frac{\gamma_{3}}{\gamma_{2}}
\alpha\beta
\left(
\begin{array}{l}
\alpha d_{l+1}(x,y)z\\
\beta d_{l+1}(x,y)z
\end{array}
\right)=
\rho_{1}
(d_{l+1}(x,y)z).
\end{array}
$$
\par
\vskip 3mm
(2)\ Case of $j=2,$
we have
\par
$$
\begin{array}{l}
[T_{l}(x,y),\rho_{2}(z)]\\
=
-\frac{\gamma_{3}}{\gamma_{2}}
\left(
\begin{array}{ll}
\alpha\beta
(d_{l+1}(x,y)+
d_{1-l}
({\bar x},{\bar y})),&
\alpha^{2}(
d_{l+1}(x,y)-
d_{1-l}
({\bar x},{\bar y})\\
\beta^{2}
(d_{l+1}(x,y)-
d_{1-l}
({\bar x},{\bar y})),&
\alpha\beta(d_{l+1}(x,y)+
d_{1-l}({\bar x},{\bar y})
\end{array}
\right)
\left(
\begin{array}{c}
k\alpha{\bar z}\\
-k\beta{\bar z}
\end{array}
\right)\\
=
-\frac{\gamma_{3}}{\gamma_{2}}
k(2\alpha\beta)
\left(\begin{array}{l}
\alpha d_{1-l}({\bar x},{\bar y}){\bar z}\\
-\beta d_{1-l}
({\bar x},{\bar y}){\bar z}
\end{array}
\right)\\
=
\rho_{2}(\overline{d_{1-l}({\bar x},{\bar y}){\bar z})}=
\rho_{2}(d_{l+2}(x,y)z).
\end{array}
$$
\par
\vskip 3mm
(3)\ Case of $j=3.$
\par
\vskip 3mm
For simplicity,
we write
$$
T_{l}(x,y)=
-\frac{\gamma_{3}}{\gamma_{2}}
\left(\begin{array}{ll}
\alpha\beta C_{l}(x,y),&
\alpha^{2}D_{l}(x,y)\\
\beta^{2}D_{l}(x,y),&
\alpha\beta C_{l}(x,y)
\end{array}
\right)$$
with
$$
\begin{array}{l}
C_{l}(x,y)=
d_{l+1}(x,y)+
d_{1-l}({\bar x},{\bar y})\\
D_{l}(x,y)=
d_{l+1}(x,y)-
d_{1-l}({\bar x},{\bar y})
\end{array}
$$
and calculate
$$
\begin{array}{l}
[T_{l}(x,y),\rho_{3}(z)]\\
=
-\frac{\gamma_{3}}{\gamma_{2}}
(k\frac{\gamma_{1}}{\gamma_{2}})
[
\left(
\begin{array}{ll}
\alpha\beta
C_{l}(x,y),&
\alpha^{2} D_{l}(x,y)\\
\beta^{2}D_{l}(x,y),&
\alpha\beta C_{l}(x,y)
\end{array}
\right),
\left(
\begin{array}{ll}
\alpha\beta l(z+{\bar z}),&
\alpha^{2}l(z-{\bar z})\\
-\beta^{2}l(z-{\bar z}),&
-\alpha\beta l(z+{\bar z})
\end{array}
\right)]\\
=
-\frac{\gamma_{3}\gamma_{1}}{(\gamma_{2})^{2}}
k
\left(
\begin{array}{ll}
\alpha^{2}\beta^{2}M_{11},&
\alpha^{3}\beta M_{12}\\
\beta^{3}\alpha M_{21},&
\alpha^{2}\beta^{2} M_{22}
\end{array}
\right)
\end{array}
$$
where we have
$$
\begin{array}{l}
M_{11}=-M_{22}=
[C_{l}(x,y),
l(z+{\bar z})]-
\{D_{l}(x,y),
l(z-{\bar z})\}_{+},\\
M_{12}=-M_{21}=
[C_{l}(x,y),
l(z-{\bar z})]-
\{D_{l}(x,y),
l(z+{\bar z})\}_{+}
\end{array}
$$
and hence
$$
\begin{array}{ll}
M_{11}&=
[d_{l+1}(x,y)-
d_{1-l}({\bar x},{\bar y}),
l(z+{\bar z})]-
\{
d_{l+1}
(x,y)-
d_{1-l}({\bar x},{\bar y}),
l(z-{\bar z})\}_{+}\\
&=
l((d_{l}(x,y)+
d_{3-l}
({\bar x},{\bar y}))
(z+{\bar z}))+
l((d_{l}(x,y)-
d_{3-l}
({\bar x},{\bar y}))
(z-{\bar z}))\\
&=
2l(d_{l}(x,y)z+
d_{3-l}
({\bar x},{\bar y}){\bar z})=
2l(d_{l}(x,y)z+
\overline{ d_{l}(x,y)z})
\end{array}
$$
and similarly
$$
M_{12}=
2l(d_{l}(x,y)z-
d_{3-l}
({\bar x},{\bar y}){\bar z})=
2l
(d_{l}(x,y)z-
\overline{ d_{l}(x,y)z}).
$$
\par
Thus, we find
$$
\begin{array}{l}
[T_{l}(x,y),
\rho_{3}(z)]\\
=
-\frac{\gamma_{1}\gamma_{3}}{(\gamma_{2})^{2}}
2k\alpha\beta
\left(
\begin{array}{ll}
\alpha\beta l(d_{l}(x,y)z+
\overline{d_{l}(x,y)z}),&
\alpha^{2}l(d_{l}(x,y)z-
\overline{d_{l}(x,y)z})
\\
-\beta^{2}l(d_{l}(x,y)z-
\overline{d_{l}(x,y)z}),&
-\alpha\beta l(d_{l}(x,y)z+
\overline{d_{l}(x,y)z})
\end{array}
\right)\\
=\rho_{3}(d_{l}(x,y)z).
\end{array}
$$
\par
This completes the proof of Eq.(3.7c).
\vskip 3mm
{\bf Proof of Eq.(3.7d)}
\par
\vskip 3mm
This follows from
Eqs.(3.7a-c)
and the fact that 
the Lie algebra associated with
structurable algebra is
contained into $\hat{L}=L(W,W)\oplus W,$
when we replace
$T_{j}(x,y)$
by
$$
T_{j}(x,y)=
\frac{\gamma_{2-j}}{\gamma_{1-j}}
[\rho_{3-j}(x),
\rho_{3-j}(y)].$$
\par
These complets the proof of Theorem 3.1.//
\par
\vskip 3mm
{\bf Concluding Remark}
\par
\vskip 3mm
Theorem 3.1 clarifies the following fact.
The Lie algebra constructed in Eqs.(3.7) 
on the basis of the structurable algebra 
is known to be a
$BC_{1}$-graded
Lie algebra of type
$B_{1}$,
provided that the ground field $F$ contain the square
root $\sqrt{-1}$
of $-1$
([E-O,2]).
On the other side,
for any left unital
$(-1,1)$FKTS
(i.e.,
we have
$eex=x$
for any
$x\in A$ where
$e$ is a left unital element),
and hence for any $A$-ternary algebra,
the associated standard Lie algebra
constructed as in Eqs.(1.13)
and (1.14)
is also a
$BC_{1}$-graded Lie algebra of type $B_{1}$
without assuming
$\sqrt{-1}\in F,$ 
(see [E-K-O]).
Also,
if $F$ is an algebraically closed field of characteristic zero,
then  
any simple Lie algebra is known to be 
$S_{4}$-invariant and can be constructed by some structurable algebra,
so that any such Lie algebra is
also a $BC_{1}$-graded Lie algebra of type $B_{1}$,
(as well as of type $C_{1}$).
Of course,
the underlying $sl(2)$ 
symmetry is different
for both $B_{1}$
and $C_{1}$
cases.
\par
\vskip 3mm
{\bf Referense}
\par
\vskip 3mm
[A]:B.N.Allison;
A class of nonassociative algebras with involution containing the
class of 
Jordan algebras. Math.Ann.,
{\bf 237} (1978)
133-156
\par
[A-B-G]:B.N.Allison, G.Benkart and Y.Gao;
Lie algebras graded by the root systems $BC_{r},\ 
r \geq 2 $, Mem. Amer. Math. Soc., vol.{\bf 158}, (2002).

\par
[A-F]:B.N.Allison and J.R.Faulkner;
Non-associative coefficient algebras for Steinberg unitary
Lie algebras. J.Algebras {\bf 161}
(1993) 1-19
\par
[B-S]:
G.Benkart and O.Smirnov;
Lie algebras graded by the
root system $BC_{1}$.
J.Lie Theory 
{\bf 13}
(2003)
91-132
\par
[E]:
A.Eldque;
The magic square and symmetric composition.
Rev.Mat.
Iberoamericana
{\bf 20}
(2004)
477-493
\par
[E-O,1]:
A.Elduque and S.Okubo;
Lie algebras with $S_{4}$-action and structurable algebras,
J.Alg., {\bf 307}
(2007)
864-890
\par
[E-O,2]:
A.Elduque and S.Okubo;
$S_{4}$-symmetry on the Tits construction of Exceptional
Lie algebras and Superalgebras.
Pub.Math.
{\bf 52}
(2008)
315-346
\par
[E-O,3]:
A.Elduque and S.Okubo;
Lie algebras with
$S_{3}$
or
$S_{4}$-action and
generalized Malcev algebras.
Proc.
Ray.Soc.
Edinb.
{\bf 139A}
(2009)
321-357
\par
[E-O,4]:
A.Elduque and S.Okubo;
Special Freudenthal-Kantor triple systems
and Lie algebras
with dicyclic symmetry.
Proc.Roy Soc.
Edinb.
{\bf 141A}
(2011)
1225-1262
\par
[E-K-O]:
A.Elduque,N.Kamiya and S.Okubo;
Left unital Kantor triple systems and
structurable algebra. preprint,
arXiv 1205.2489.
(2012) 
\par
[Ka]:
N.Kamiya;
A construction of simple Lie algebras over 
$C$ from balanced Freudenthal-Kantor
triple systems.
Contribute to general algebras,
{\bf 7},
(Verlag Holder-Pickler-
Tempsky,
Wien )
(1991) 205-213.
\par
[K-N]:
S.Kobayashi and K.Nomizu;
Foundation of Differential Geometry,
I, 1963; II 1968, Wiley-Interscience, New York.
\par
[Ko]:
M.Kochetev;
Gradings on finite-dimensional simple
Lie algebras.
Acta Appl.
Math.
{\bf 108}
(2009)
101-129
\par
[K-O]:
N.Kamiya and S.Okubo;
On $\delta$-Lie Supertriple systems Associated with
$(\varepsilon,\delta)$ Freudenthal-Kantor Triple Systems.
Proc.
Edinb.
Math.
Soc.
{\bf 43}(2000)
243-260
\par
[K-M-O]:
N.Kamiya,D.Mondoc,
and S.Okubo;
A structure Theory of
$(-1,-1)$ Freudenthal-Kantor triple systems.
Bull.Aust.
Math.Soc.
{\bf 81}
(2010)
132-155
\par
[M]:
K.Meyberg;
Eine Theorie der Freudenthalschen
Triple systeme,I,II.
Nederl. Acad. Wetensch. Ser.
A-{\bf 71}
=Indag Math.
{30}
(1968)
162-174 and 175-190
\par
[O]:
S.Okubo;
Symmetric triality relations and structurable
algebra.
Linear Algebra and its Application
{\bf 396}
(2005)
189-222
\par
[Y-O]:
K.Yamaguti and S.Ono;
On representations of
Freudenthal-Kantor triple system
$ U(\varepsilon,\delta)$.
Bull.Fac.
School,Ed.
Hiroshima Univ.
Ser.II,
{\bf 7}
(1984)
43-51
\end{document}